\documentclass[usenatbib]{mn2e}
\usepackage{graphicx}
\usepackage{times}
\usepackage{mathptmx}
\usepackage{epsfig}

\begin{document}

\title{[OII] emitters in the GOODS field at \emph{z}$\sim1.85$: a homogeneous measure of evolving star formation.}
\author[K.D.Bayliss]{K.D.Bayliss$^{1}$\thanks{E-mail: kdb25@ast.cam.ac.uk}, R.G.McMahon$^{1}$, B.P.Venemans$^{2}$, E.V. Ryan-Weber$^{3}$, J.R.Lewis$^{1}$\\
$^{1}$Institute of Astronomy, Madingley Road, Cambridge CB3 0HA\\
$^{2}$European Southern Observatory, Karl-Schwarzschild Strasse, 85748 Garching bei M\"{u}nchen, Germany\\
$^{3}$Centre for Astrophysics \& Supercomputing, Swinburne University of Technology, Mail H39, PO Box 218, Hawthorn, 3122 VIC, Australia}
\maketitle

\begin{abstract}
We present the results of a deep, near-infrared, narrow band imaging survey at a central wavelength of 1.062 $\mu$m (FWHM=0.01$\mu$m) in the GOODS-South field using the ESO VLT instrument, HAWK-I. The data are used to carry out the highest redshift search for [OII]$\lambda$3727\AA $\;$emission line galaxies to date. The images reach an emission line flux limit (5$\sigma$) of 1.5 $\times$ $\rm 10^{-17} erg \; cm^{-2}\; s^{-1}$, additionally making the survey the deepest of its kind at high redshift. In this paper we identify a sample of [OII]$\lambda$3727\AA $\;$emission line objects at redshift \emph{z}$\sim$1.85 in a co-moving volume of $\rm \sim4100\; Mpc^{3}$. Objects are selected using an observed equivalent width (EW$_{obs}$) threshold of EW$_{obs}>$ 50\AA. The sample is used to derive the space density and constrain the luminosity function of [OII] emitters at \emph{z}=1.85. We find that the space density ($\rho$) of objects with observed [OII] luminosities in the range $\rm log(L_{[OII]}) > 41.74 \; erg\; s^{-1}$ is log($\rho$)=$-$2.45$\pm0.14\;$Mpc$^{-3}$, a factor of 2 greater than the observed space density of [OII] emitters reported at \emph{z}$\sim$1.4. After accounting for completeness and assuming an internal extinction correction of A$_{H\alpha}$=1 mag (equivalent to A$_{[OII]}$=1.87), we report a star formation rate density of $\dot{\rho}_* \sim$0.38$\pm0.06$ M$_\odot$ yr$^{-1}$ Mpc $^{-3}$. We independently derive the dust extinction of the sample using 24$\mu$m fluxes and find a mean extinction of A$_{[OII]}$=0.98$\pm0.11$ magnitudes (A$_{H\alpha}$=0.52). This is significantly lower than the A$_{H\alpha}$=1 (A$_{[OII]}$=1.86) mag value widely used in the literature. Finally we incorporate this improved extinction correction into the star formation rate density measurement and report $\dot{\rho}_* \sim0.24\pm0.06$ M$_\odot$ yr$^{-1}$ Mpc $^{-3}$.

\end{abstract}


\section{Introduction} 

The volume-averaged star formation history is a fundamental property of the universe: its reliable determination will provide a powerful probe with which to explore the physics of galaxy formation and evolution. 

Much effort has been focused in this field and a variety of star formation rate (SFR) indicators at different redshifts have contributed to the current picture whereby star formation starts between \emph{z}$\sim$20-8, peaks at \emph{z}$\sim$1-3 and then declines towards  \emph{z}=0. Although constrained within $30$-50\% below \emph{z}$\sim$1, the star formation rate is less well determined at higher redshifts, being uncertain up to a factor of 3 between $1<z<6$ (Hopkins \& Beacom, 2006). \nocite{hop06}

Variation between different measurements is primarily due to differences in sample selection, biases between different SFR indicators and underlying cosmic variance. Whilst using a combination of different indicators has provided a qualitative description of the evolution of the SFR, such biases make it difficult to properly quantify the evolution. Indeed, as pointed out by \cite{geach08}, piecing together measurements from different indicators is no longer improving our understanding.

To make progress, homogeneous indicators are needed, visible over wide redshift ranges. Although H$\alpha$ remains the SFR indicator of choice, the [OII]$\lambda$3727 doublet has a particular advantage over H$\alpha$ in being visible to \emph{z}$\sim 5$ in the near-infrared, compared to the \emph{z}$\sim 2.5$ limit of H$\alpha$ surveys.

Up until recently, [OII] has for the most part remained on the periphery of efforts to measure the SFR history of the universe. As a collisionally excited forbidden line, the [OII] doublet is not directly coupled to the UV ionising radiation and as such, the [OII]-SFR calibration is subject to scatter due to excitation variations related to metal abundances and ionisation state (see the review by Kennicutt, 1998\nocite{ken98}). Despite these limitations, the intrinsic [OII]/H$\alpha$ variation is typically a factor of two or less over wide ranges of galaxy environments and abundances and [OII] has been employed in a range of previous studies (see Hippelein et al. 2003; Hopkins 2004; Kewley, Geller \& Jansen 2004; Ly et al. 2007; Takahashi et al. 2007; Zhu, Moustakas \& Blanton 2009) \nocite{hop04,hip03,ly07,tak07,zhu09} acting as a useful index, particularly for large statistical samples.

Recent advances in the calibration of the [OII]$\lambda 3727$ doublet (see Kewley et al. 2004; Moustakas, Kennicutt \& Tremonti 2006; Kennicutt et al. 2009)\nocite{kew04,mou06,ken09} have made it possible to use [OII] with greatly improved precision, approaching that of more traditional SFR indicators. Addressing concerns over metallicity and excitation, Moustakas et al. (2006) found that the majority of variation between [OII] and SFR is due to dust-reddening (derived from the H$\alpha$/H$\beta$ decrement) and that variations in metallicity and excitation are in fact second-order effects in most galaxies.

Taking this further, Kennicutt et al. (2009) developed empirical calibrations between [OII] luminosity and SFR using weighted combinations of either Total Infrared (TIR), 24$\mu$m or 8$\mu$m flux to correct for dust extinction. They report that for \emph{z}=0 galaxies, the dispersion of [OII] flux, corrected using their empirical relations, is equivalent to that of their corrected H$\alpha$ samples, thus facilitating the first reliable [OII]-derived measurements of the SFR to be made.

In this paper, we apply these advancements in the calibration of [OII] to the highest redshift narrow band survey for [OII] emitters to date, concentrating on objects in the GOODS field at \emph{z}=1.85. Using the new HAWK-I instrument on the ESO VLT facility, the survey covers a co-moving volume of $\sim 4100$Mpc$^3$ to a depth of 1.5 $\times$ $\rm 10^{-17} erg \ cm^{-2}\; s^{-1}$.

In Section 2 we describe the data set, data reduction and cataloguing techniques used in the study. Section 3 describes the method for selecting emission line galaxies (ELGs) and how we remove emitters other than [OII] from the sample. After taking into account the completeness, we compute the luminosity function in Section 4. Section 5 looks at the number density evolution of [OII] emitters between \emph{z}=0.8 and \emph{z}=1.85. In Section 6, we convert the integrated [OII] luminosity into a star formation rate, firstly using a standard extinction correction and secondly using the improved locally derived calibrations of Kennicutt et al. (2009).
Throughout this paper, magnitude measurements are on the AB scale (m$_{AB}=48.60-2.5$log$_{10}$flux). A standard cosmology is assumed with $\Omega_M=0.3, \Omega_\lambda=0.7$ and $h=0.70.$

\section{Observations and Data Reduction}
\label{obsanddatred}
\subsection{Observations}
\begin{figure}
\begin{center}
\epsfig{file=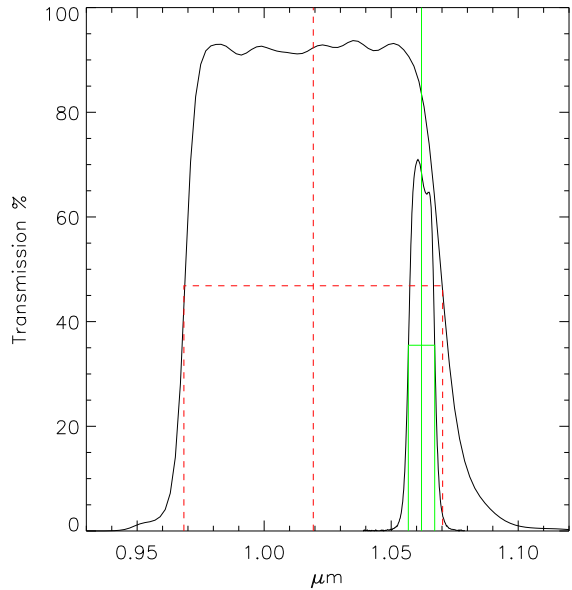,width=1.0\linewidth,clip=}
\end{center}
\caption{Transmission curves of the Y and NB1060 filters used in this
  study. Filter centres and FWHM values are indicated by dashed and
  solid lines for the Y and NB1060 filters respectively.}
\label{filters}
\end{figure}

This study utilises deep NIR data in two overlapping bands, a broad Y
filter centred on 1.021$\mu$m and a narrow band filter at
1.060$\mu$m. Filter centres and FWHM values were calculated using the
prescription in Pascual, Gallego \& Zamorano
(2007)\nocite{pascual07}. Filter transmission profiles are plotted in
Figure \ref{filters}. The data were obtained using HAWK-I
(Kissler-Patig et al. 2008\nocite{kissler08}), a NIR (0.85-2.5$\mu$m)
wide field, cryogenic imager on UT4 at the ESO VLT facility in
Paranal, Chile. The HAWK-I focal plane is made up of 4 square Hawaii
detectors separated by a cross-shaped gap of 15$^{\prime\prime}$. Each
filter has a 2048 pixel width and a pixel scale of
0.106$^{\prime\prime}$/pix. The HAWK-I field of view of is 7.5$\times$
7.5 arcmin. The data were collected in the Science Verification phase
as part of program 60.A-9284(B) \emph{Fontana et al.: A deep infrared
  view on galaxies in the early Universe.} A single pointing was taken
in a region of the GOODS field centred on co-ordinates 3h 32m 29.0s,
-27$^o$44$^{\prime}$28$^{\prime\prime}$.

Table \ref{filterinfo} gives basic filter information, the average
seeing of the images and the total exposure time in each filter.

\begin{table}
\centering
\begin{minipage}[l]{0.5\textwidth}
\caption{Filter Statistics and Image Seeing.}
  \begin{tabular}{lllll}
 \hline
Filter    & $\lambda_c$ ($\mu$m) & $\Delta \lambda$ ($\mu$m) & Seeing ($^{\prime\prime}$) & Exposure Time\\
&&&&                                                                             (hr:min:s)\\
 \hline
NB1060    & 1.0619       &  0.0104    &   0.74   & 8:20:00  \\
Y         & 1.0193       &  0.1018   &   0.57   & 1:10:30  \\
  \hline
  \end{tabular}
  \label{filterinfo}
  \end{minipage}
\end{table}

\subsection{Reduction and calibration}
The observations are made up of a series of spatially offset
(dithered) exposures, comprising $141\times 30$s exposures in Y and $100\times300$s exposures in NB1060.
The data was reduced using a pipeline specially developed for HAWK-I at the 
Cambridge Astronomy Survey Unit that incorporates components of the
VISTA Data Flow System (Irwin et al., 2004)\nocite{irwin04}. The pipeline can be summarised as follows:

Firstly, each exposure has its dark current and flat field instrumental signature removed.  
The dark current was subtracted using a master dark image created from
dark frames with the same exposure time as the image (about 100 dark frames were available per exposure time).

The images were then flat fielded to remove the pixel to pixel quantum
efficiency variation as well as the large scale vignetting profile.  A
master flat field for each filter was formed from exposures of the
twilight sky, which were scaled to bring them to a common median
background and then combined using a mean combination algorithm with sigma-clipping (5$\sigma$). The telescope was moved
slightly between each twilight flat exposure so that when twilight
flat field images are combined, any remaining astronomical objects
are removed by the rejection algorithm.  

Any pedestal scale factors between the
individual detectors (due to differences in gain or average
QE) were removed by normalising the images by the ensemble median of the background flux.

Once the jittered exposures are dark and flat corrected, they are
median combined to form a sky background image. The background images
are normalised to a zero median so that the median background level in
the corrected images is preserved.  The background corrected images
from each jitter sequence are registered internally using
visible sources and then shifted and combined.  

A world coordinate system is added by fitting sources detected on the
image to the 2MASS point-source catalogue. Finally, all the
stacks are combined together to form a single deep stack for each
filter.

The zero point of the Y images was determined by calibrating stellar
counts onto ISAAC J and H images (Retzlaff et
al. 2010)\nocite{retzlaff10}, using synthetic $J-H$ and $Y-J$ colours
generated by \cite{hewett06} as reference. $Y-NB1060$ colours
generated by the same synthetic codes were used to calibrate the
NB1060 images onto the Y band.

The images were then scaled to a common zero point of 30.0 (AB), accurate to 0.1 magnitudes. 

\subsection{Cataloguing and photometry}
\label{cataloguing}

The NB1060 and Y images are registered using stellar point sources as
a reference. Similarly, the images are PSF matched by smoothing the Y
image with a gaussian kernel until the stellar FWHM values are equal
to those measured in the NB1060 image. Source Extractor (Bertin \&
Arnouts, 1996\nocite{bertin}) is then run on the NB1060 image to
produce a NB1060 selected source catalogue. Extraction parameters are
given in Table \ref{separams}. We find slight deviations in the
background of the order of a few percent in the NB1060 detector 2
image. This appears to be due to radio active decay tracks created
near to the detector. For this reason, for this detector only, the
background is determined by filtering over smaller areas to produce a
more accurate local background for each object. The altered parameters
are given in the final column of Table \ref{separams}. Values were
chosen to minimise the fluctuations in the background measured by
Source Extractor (assessed by outputting a `check image' of type
`BACKGROUND')

\begin{table}
\centering
\begin{minipage}[l]{0.5\textwidth}
\caption{Source Extractor parameters - NB1060 source catalogue.}
  \begin{tabular}{llll}
\hline
 Parameter        & Unit       & Value\footnote{Parameters used for detectors 1, 3 and 4} & Value\footnote{Changes made to standard parameter input for detector 2 to compensate for low-level background variations}    \\
  \hline  
 DETECT\_MINAREA  & pix        & 8      &    \\
 DETECT\_THRESH   & $\sigma$   & 1.5    &    \\
 ANALYSIS\_THRESH & $\sigma$   & 1.5    &    \\
 BACK\_SIZE       & pix        & 64     &  15\\
 BACK\_FILTERSIZE & pix        & 5      &  3 \\
 DEBLEND\_NTHRESH & ADU        & 32     &    \\
 CLEAN            &            & Y      &    \\
 CLEAN\_PARAM     &            & 2.0    &    \\
 PIXEL\_SCALE     & $^{\prime\prime}$/pix     & 0.1064 &    \\
  \hline
  \end{tabular}
  \label{separams}
  \end{minipage}
\end{table}

For each object in the catalogue, two flux measurements are made: $Y-NB1060$ colour and total NB1060 flux.
Circular apertures are used throughout. The size of each aperture is tuned to the size of the object as described in Labbe et al. (2003)\nocite{labbe03}\footnote{We note that if Source Extractor's elliptical `AUTO' aperture is used in place of the TOT aperture presented here, this changes the final SFR measurements by $< 3\%$.}. The Labbe et al. scheme recommends different sized apertures for making colour and total flux measurements, including modifications for blended, extended and particularly compact objects. Colour measurements are made in circular apertures of diameter D$_{COL}$=2(A$_{ISO}/\pi)^{1/2}$, where A$_{ISO}$ is the measured isophotal area within the detection isophote. Similarly, total flux measurements are made in apertures of diameter D$_{TOT}$=2(A$_{KRON}/\pi)^{1/2}$ where A$_{KRON}$ is the Kron area; the area of the ellipse defined by the Kron radius \citep[see][]{kron} (in Source Extractor this is the area of the `AUTO' aperture).

For colour measurements, the object is detected in the narrow band and then equal sized circular apertures are placed in the same position on both the NB1060 and Y images, in an analogous way to using Source Extractor in dual image mode. Colour measurements therefore have the same spatial origin.

Objects within $\sim10^{\prime\prime}$ of the edges of the stacks are removed due to the lower exposure time in these regions, leaving a survey area of 46.24 square arcmin. Detections with a S/N $< 3.0$ in the narrow band are additionally removed from the catalogue.

There are 2150 objects in the full catalogue. 
We find the 5$\sigma$ NB1060 flux limit in a circular aperture, 10 pixels (1.06$^{\prime \prime}$) in diameter is 1.5 $\times$ $\rm 10^{-17} erg \ cm^{-2} s^{-1}$ (m$_{NB1060}$=24.55). 
To estimate the point source completion of the catalogue, we insert artificial point sources into the images using the IRAF program MKOBJECT and extract them using Source Extractor as described above. We find the 90\% completeness limit of point sources in the NB1060 catalogue is m$_{NB1060}$=24.40 (1.73 $\times$ $\rm 10^{-17} erg \ cm^{-2} s^{-1}$.)

\section{Emission Line Galaxy Selection}
\label{elgselection}
We expect emission line galaxies (ELGs) to have excess NB1060 flux compared to the Y band continuum.
ELGs are therefore selected based on a clear flux excess in the narrow band, ie, $Y-NB1060$ $>$ 0.
An object is selected based on two criteria:

The parameter $\Sigma$ \citep{bunker95} is used to characterise the significance of the NB1060 excess compared to a flat spectrum, taking into account the noise properties of the images.
For this work, the appropriate selection curve is given by

\begin{equation}
m_{Y}-m_{NB} = -2.5\log_{10} \left[1-\Sigma 10^{-0.4(30.0-m_{NB})}\sqrt{\sigma_{NB}^2+\sigma_Y^2} \right]
\end{equation}
where $\sigma_{NB}$ and $\sigma_Y$ are the noise in the NB1060 and Y images respectively and 30.0 is the scaled zero-point as described in Section \ref{cataloguing}.

We use a colour significance of $\Sigma$=3 to select the ELG candidates (\cite{bunker95}). This assures that the fraction of ordinary, non emission line galaxies scattered into the sample due to noise is very low,  $\sim$ 1 in 1000 objects.

Secondly an observed equivalent width (EW$_{obs}$) criterion of EW $>$ 50\AA $\;$is imposed, equivalent to a colour cut of 
\begin{equation}
Y-NB1060 > 0.37.
\end{equation}
This corresponds to a rest-frame equivalent width of 17.5\AA $\;$ for [OII] at \emph{z}=1.85.

\begin{figure}
\begin{center}
\epsfig{file=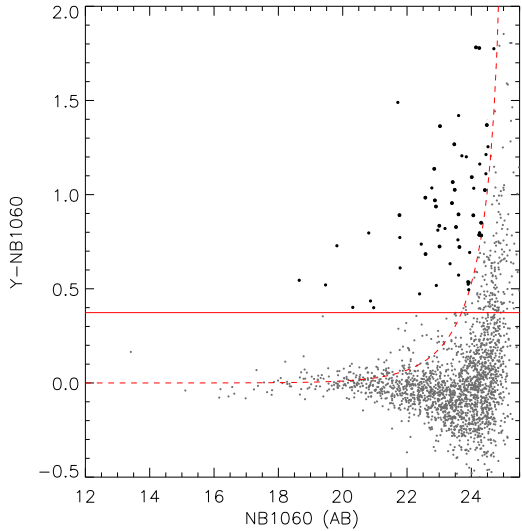,width=0.9\linewidth,clip=}
\end{center}
\caption{Colour-magnitude candidate selection diagram. The dashed red line shows the $\Sigma$=3 selection criterion and the solid red line indicates the 50\AA$\;$ observed (17.5\AA $\;$rest frame) equivalent width cutoff. ELG candidates are highlighted in black.}
\label{selection}
\end{figure}

Figure \ref{selection} shows the colour magnitude selection diagram along with the $\Sigma$=3.0 selection line (curved) and the equivalent width criterion (solid horizontal line). Candidate ELGs are highlighted in black. Using these criteria, 58 objects are selected as ELG candidates. 

\subsection{Selection of [OII] emitters at \emph{z}=1.85}

The 58 NB selected ELGs are expected to be comprised of [OII]$\lambda3727$,  H$\alpha$, [OIII]$\lambda\lambda$4959,5007 and H$\beta$ emitters, along with a small fraction of redshift interlopers. Possible emission lines and their corresponding redshifts and volumes are given in Table \ref{elginfo}.

\begin{table}
\centering
\begin{minipage}[l]{0.5\textwidth}
\caption{Emission Lines Possibly Detected in the NB1060 filter.}
  \begin{tabular}{lllllll}
 \hline
 Emission & $\lambda_{rest}$ & \emph{z} & dz & V$_{c}$ \footnote{Co-moving volume defined by the NB1060 filter width given in Table \ref{filterinfo}.} \
                                     & log(L$_{Lim}$) \footnote{Luminosity corresponding to a NB1060 line flux of $\log(f_{line})= -16.82\;$erg s$^{-1}$ cm $^{-2}$, the 5$\sigma$ NB1060 limit of the data (1$^{\prime \prime}$ aperture)}\
                                     &  SFR$_{Lim}$  \footnote{Limiting SFR corresponding to the limiting luminosity in column 6, assuming SFR=7.9x10$^{-42}$ L$_{H\alpha}$ (Kennicutt 1998) and [OII]/H$\alpha$=0.45 (see Section \ref{commonobscorr}.)}\\

 Line     & ($\mu$m)        &   &    &  (Mpc$^{-3}$)   & erg s$^{-1}$ & M$_\odot$ yr$^{-1}$\\
 \hline
 H$\alpha$      & 0.6563 & 0.62 & 0.016 & 950   & 40.38 &  0.19\\
 {[}OIII]       & 0.5007 & 1.12 & 0.021 & 2353  & 41.02 & \\
 {[}OIII]       & 0.4959 & 1.14 & 0.021 & 2410  & 41.04 & \\
 H$\beta$       & 0.4861 & 1.18 & 0.021 & 2528  & 41.08 & \\
 {[}OII]        & 0.3727 & 1.85 & 0.028 & 4138  & 41.56 & 6.34\\
 Ly$\alpha$     & 0.1216 & 7.73 & 0.085 & 7977  & 43.04 & \\
 \hline 
  \end{tabular}
  \label{elginfo}
  \end{minipage}
\end{table}

We split the sample into composite line samples in two ways:
(1) We use the spectroscopic and photometric redshifts reported in the GOODS MUSIC catalogue, version 2 (Santini et al., 2009\nocite{santini09}).  
(2) We use galaxy evolution tracks in the colour-colour diagram $z-J$, $V-i$ to split the ELG sample into low and high redshift populations according to colour.

Photometric redshifts in the MUSIC catalogue are based on photometry in 15 bands spanning the optical to the far infrared. We find good agreement between the 1697 spectroscopic redshift measurements in the catalogue and their equivalent photometric redshift estimates. Splitting the MUSIC measurements into low ($z_{phot}< 1.5$) and high ($z_{phot}> 1.5$) redshift groups, we find the median offset of the low redshift group is $\langle \emph{z}_{phot}-z_{spec} \rangle = -0.004$ with a dispersion of $\sigma_{MAD}=0.072$ (where $\sigma_{MAD}$ is the standard deviation associated with the median absolute deviation of $\langle \emph{z}_{phot}-z_{spec} \rangle$) and equivalently for the high redshift group, $\langle \emph{z}_{phot}-z_{spec}\rangle =+0.05$ with a dispersion of $\sigma_{MAD}=0.18$. 

\begin{figure}
\begin{center}
\epsfig{file=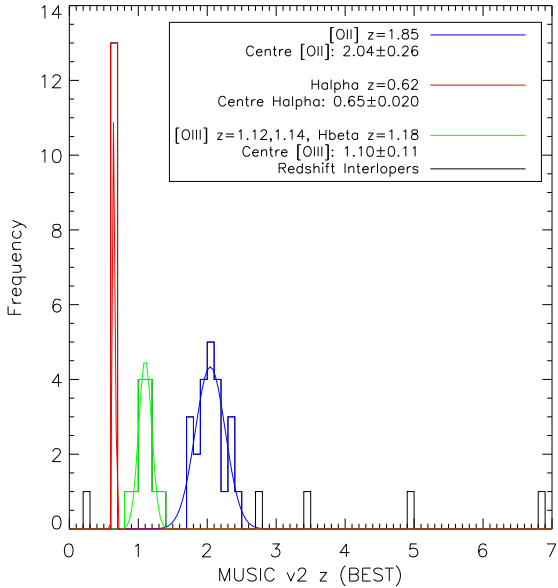,width=1.0\linewidth,clip=}
\end{center}
\caption{MUSIC catalogue photometric redshifts for the sample of ELGs, showing three peaks corresponding to [OII] emitters (blue), H$\alpha$ emitters (red) and a combination of [OIII] and H$\beta$ emitters (green). Interloping galaxies are shown in black.}
\label{zhist}
\end{figure}

Our catalogue is matched to GOODS MUSIC using a search radius of (0.5$^{\prime \prime}$). All but two candidates have MUSIC matches. 
Figure \ref{zhist} shows a histogram of the photometric redshift measurements (ZPHOT parameter in MUSIC) of candidate ELG galaxies. The histogram clearly shows 3 peaks corresponding to [OII]$\lambda$3727 at $z_{phot}\sim 2$, H$\alpha$ at $z_{phot} \sim 0.6$, and a merged peak at $z_{phot} \sim 1$ comprised of [OIII]$\lambda\lambda$ 4959,5007 and H$\beta$ emitters. 
We discard 5 galaxies with \emph{z}$_{phot}<0.3$ or  \emph{z}$_{phot}>2.7$ as redshift interlopers ($\sim$8\% of the sample), noting that one with a photometric redshift of 6.88 is likely to be a T-dwarf \citep{eyles07}.
Gaussian curves are fitted individually to each population. Fit parameters are $z_{phot,[OII]}=2.05 \pm 0.27$ for the [OII] candidate population and $z_{phot,H\alpha}=0.62\pm 0.032$ for the H$\alpha$ population. The broader spread of the [OII] peak compared to the H$\alpha$ reflects the reduced accuracy of the photometric redshifts at high redshift.

For the second diagnostic, we calculate galaxy evolution tracks in {\bf $z-J$}, {\bf $V-i$} colour space using the photometric redshift code detailed in \cite{banerji10}. Galaxy types E, Sbc and Scd are modelled using the average observed spectra of Coleman, Wu \& Weedman (1980)\nocite{coleman80}. In addition we use the observed starburst model (SB2) from \cite{kinney96} and a synthetic spectrum of a galaxy with an instantaneous 50Myr star burst, generated using the \scriptsize PEGASE \normalsize code \citep{fioc97}. 

Each model is redshifted in steps of 0.01 producing tracks from \emph{z}=0.6 (the lowest redshift, H$\alpha$ emitters) to \emph{z}=1.85 (the highest redshift [OII]$\lambda$3727 emitters). 

\begin{figure}
\begin{center}
\epsfig{file=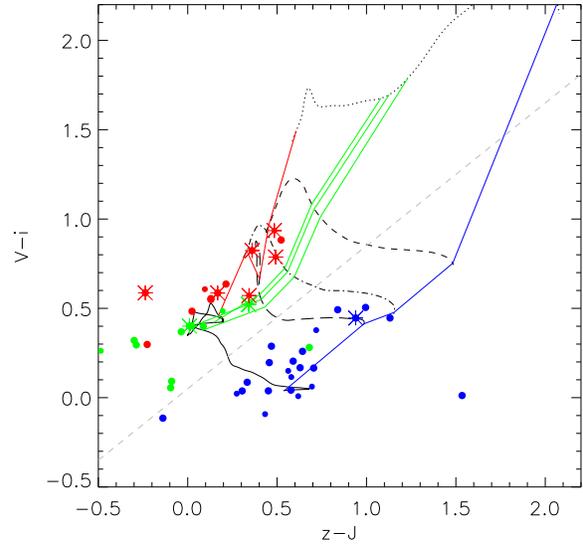,width=1.0\linewidth,clip=}
\end{center}
\caption{Splitting ELGs into low and high redshift groups according to position in {\bf $z-J$}, {\bf $V-i$} space. Black lines show galaxy evolution tracks of different galaxy models redshifted from \emph{z}=0.6-z=1.85. Top to bottom these are: E, Sbc, Scd, 50Myr instantaneous starburst and Star Burst model SB2 (See text for details). The blue line joins up points on the tracks corresponding to models at \emph{z}=1.85, indicating where [OII] emitters are expected to lie in the diagram. Similarly, the red line joins points at \emph{z}=0.6, showing where H$\alpha$ emitters are expected to lie and the green lines join points at \emph{z}=1.12,1.14 and 1.18 corresponding to [OIII]$\lambda\lambda$4959,5007 and H$\beta$. Candidate emitters are colour coded according to their photometric redshifts as in Figure \ref{zhist} with [OII] candidates in blue, [OIII] and H$\beta$ in green and H$\alpha$ in red. Asterisks indicate spectroscopically confirmed candidates (Santini et al. 2009).} 
\label{tracks}
\end{figure}

In Figure \ref{tracks}, the evolutionary tracks are indicated by black lines, where (from top to bottom) the elliptical track is shown in the dotted line, type Sbc in dash, Scd in dash-dot, the 50Myr starburst model in long-dash and the starburst (SB2) model in the solid line. Joining up the points on the five tracks at the redshift of each line emitter in the filter delineates where each population of line emitters is expected to lie in colour-colour space. The blue line joins points at \emph{z}=1.85, indicating where [OII]$\lambda$3727 emitters are expected to lie. Similarly, the red line joins points at \emph{z}=0.63 corresponding to H$\alpha$ and lines joining points at \emph{z}=1.12,1.14 and 1.18 are shown in green corresponding to [OIII]$\lambda\lambda$4959,5007 and H$\beta$ (top to bottom respectively). 

We use {\bf $z-J$}, {\bf $V-i$} photometry from the MUSIC catalogue to over-plot the ELGs. The ELGs in the sample, indicated by circles, are blue in {\bf $z-J$}, {\bf $V-i$}; consistent with them being star forming galaxies. The candidates clearly split into two regions in colour space, the lower stream attributed to [OII] objects and the upper to a combination of H$\alpha$, H$\beta$ and [OIII] emitters. 

Final classifications, taking both colour and photometric redshift into account, are colour coded on the Figure with objects classified as [OII] shown in blue (26 objects), H$\alpha$ in red (14 objects) and H$\beta$ or [OIII] objects in green (13 objects). One object has a photometric redshift placing it as an [OIII] or H$\beta$ emitter, yet appears in the [OII] stream in Figure \ref{tracks}. Upon visual inspection, this object appears to be blended with a neighbouring object in the image which may be affecting the measurements and we therefore discard it from the sample. Ten objects have good quality spectroscopic redshifts and these are indicated in the Figure with asterisks. 

Over all, there is good agreement between the track, photometric redshift and spectroscopic classifications. When the \emph{z}$_{phot}$ classification is taken into consideration, objects in Figure \ref{tracks} clearly split into three classes, with [OIII] and H$\beta$ emitters lying bluer than H$\alpha$ in {\bf $V-i$}.

Four objects are left unclassified: two with no MUSIC match and a further two with partial entries in MUSIC, not including JHK or photometric redshift measurements. These are classified using $B-i$, $V-z$ colours in the same manner as in Figure \ref{tracks}. For the two objects with no MUSIC data, colours are determined from ACS cutouts generated by the MAST cutout tool and the documented zero-points for the ACS images. On the basis of the positions of these objects in $B-i$, $V-z$ space, three were classified as [OII] emitters and the remaining object was classified as either an [OIII] or H$\beta$  emitter.

We checked the sample was uncontaminated by stars by visually inspecting each object and additionally ensuring that none of the objects were flagged as stars in the MUSIC catalogue. None of the objects were flagged as AGN in MUSIC and none of our objects matched the Chandra Deep Field-South: 2Ms Source Catalogue (Luo et al. 2008\nocite{luo08}), suggesting that the sample is additionally uncontaminated by AGN.

In summary, of 58 initial ELG candidates, 53 were identified as genuine ELGs. Our final [OII] emitter sample contains 26 objects. 

\section{The Observed Luminosity Function}
\label{sfr1}

The observed luminosity function is calculated in 4 stages as follows:
In Section \ref{const_vol_lf}, the luminosity function is computed under the simplifying assumption that the NB1060 filter is a perfect top hat function with a width equal to the FWHM of the real filter. 
In Section \ref{completeness_lim} we investigate the completeness limit of the survey. 
In Section \ref{sfit} we fit a Schechter function to the complete region of the luminosity function and finally in Section \ref{filter_corr}, simulations are undertaken to scale the luminosity function to take into account the real filter shape and produce the corrected observed luminosity function.

\subsection{Line fluxes and [OII] luminosities.}

The narrow band contains both line and continuum emission. We correct for continuum emission using the Y flux density. Given that the NB1060 and Y bands overlap, the appropriate equation for the line flux is given by:

\begin{equation}
f_l= \frac{f_{NB1060}-\epsilon f_Y}{1-\epsilon}
\end{equation}
where $f_{NB1060}$ and $f_Y$ are the total fluxes in the narrow and broad band filters and $\epsilon$ is the ratio of the widths of the narrow and broadband filters (in this survey $\epsilon$=0.102).

Assuming all objects in the [OII] sample lie at the centre of the filter at \emph{z}=1.85 and the luminosity distance, d$_L$ = 4.36 $\times 10^{28}$ cm, the observed [OII] line luminosities are: 

\begin{equation}
  L[OII]_{obs}=4 \pi d_L^2 \: f_l.
\end{equation}

NB1060 and Y magnitudes, along with the line fluxes and line luminosities for the sample of 26 [OII] emitters are given in Table \ref{table}.

\begin{table}
\centering
\begin{minipage}[l]{0.45\textwidth}
\caption{Properties of the [OII] Sample.}
\setlength{\tabcolsep}{5pt}
\fontsize{7}{10}\selectfont

\begin{tabular}{ c c c c c c c c }
 R.A\footnote{Measurements in degrees} & DEC{\normalsize $^a$}& NB\footnote{NB1060 and Y magnitudes are on the AB scale}& $\sigma_{NB}$\footnote{Photometric Error}& Y{\normalsize $^b$}& $\sigma_Y${\normalsize $^c$} & {\scriptsize $\log(f_L)$}\footnote{Line Flux ($erg \; cm^{-2} s^{-1}$)}& {\scriptsize $\log(L[OII]_{obs})$}\footnote{Observed Line Luminosity ($erg \; s^{-1}$)}\\
 (J2000)  &  (J2000) & & & & \\
\hline
53.15124&  -27.79239  &23.02&0.05&24.39&0.08&-16.32&42.06\\
53.15497&  -27.79037  &23.52&0.08&24.35&0.09&-16.67&41.71\\
53.16162&  -27.78743  &22.58&0.04&23.26&0.04&-16.36&42.02\\
53.15561&  -27.77930  &21.77&0.02&22.66&0.03&-15.94&42.44\\
53.15340&  -27.78093  &24.42&0.14&25.44&0.17&-16.96&41.42\\
53.15446&  -27.77971  &22.85&0.04&23.98&0.06&-16.30&42.08\\
53.15278&  -27.78011  &24.31&0.13&25.09&0.15&-17.00&41.38\\
53.14917&  -27.77876  &23.00&0.05&23.84&0.06&-16.46&41.92\\
53.15284&  -27.77247  &23.60&0.07&24.50&0.08&-16.67&41.71\\
53.15600&  -27.77086  &24.48&0.14&25.85&0.21&-16.90&41.47\\
53.15227&  -27.77005  &22.57&0.03&23.55&0.05&-16.23&42.15\\
53.19048&  -27.75691  &23.63&0.08&24.35&0.08&-16.76&41.62\\
53.16382&  -27.76531  &23.47&0.07&24.74&0.11&-16.52&41.86\\
53.09378&  -27.81859  &24.14&0.17&25.93&0.35&-16.71&41.67\\
53.08962&  -27.77184  &23.40&0.06&24.35&0.07&-16.57&41.81\\
53.08855&  -27.76735  &23.42&0.06&24.49&0.08&-16.55&41.83\\
53.12508&  -27.76784  &24.25&0.11&26.02&0.22&-16.75&41.63\\
53.18494&  -27.71257  &23.91&0.12&24.45&0.11&-16.99&41.39\\
53.18349&  -27.71222  &24.24&0.13&25.03&0.14&-16.97&41.40\\
53.12646&  -27.69629  &24.01&0.10&25.11&0.12&-16.78&41.60\\
53.13187&  -27.68209  &24.30&0.12&25.15&0.14&-16.97&41.41\\
53.17361&  -27.70122  &23.48&0.07&24.51&0.08&-16.58&41.80\\
53.05476&  -27.75030  &22.87&0.04&23.84&0.06&-16.35&42.02\\
53.06393&  -27.74213  &23.01&0.04&23.74&0.05&-16.51&41.87\\
53.09949&  -27.70320  &24.06&0.09&24.95&0.12&-16.86&41.52\\
53.09477&  -27.70320  &22.90&0.04&23.84&0.05&-16.38&42.00\\ 

\end{tabular}
\label{table}
\end{minipage}
\end{table}

\subsection{Fixed-volume luminosity function}
\label{const_vol_lf}

The luminosity function is initially calculated by assuming the NB1060 filter is a perfect top hat function. Under this approximation, all objects are visible through the FWHM of the filter, leading to a constant survey volume. 
Additionally, we assume the filter is sufficiently narrow such that it has uniform sensitivity to line strengths throughout the full filter width (ie. moving an object of fixed intrinsic line luminosity across the redshift range defined by the filter will not significantly alter its observed line flux.) 

For clarity, the resulting binned luminosity function is referred to as the `fixed-volume' luminosity function, given by the relation   

\begin{equation}
\phi_i(\log L([OII]))=\frac{1}{\Delta(\log L([OII]))} \frac{N_{i}}{V_c}
\label{calcphi}
\end{equation}
where V$_c$ is the fixed co-moving volume probed by the filter, $\Delta$(log L([OII])) is the bin width and N$_i$ is the number of galaxies with [OII] luminosity in the range log L([OII])$\pm$ 0.5$\Delta$log L([OII]). When assuming a top hat filter function, V$_c$ is fixed to the volume defined by the FWHM of the NB1060 filter (0.0104 $\mu$m.) With a survey area of 46.24 square arc minutes, at \emph{z}=1.85 the survey covers a co-moving volume of V$_c$=4138 Mpc$^{-3}$. 
The fixed-volume luminosity function is tabulated in Table \ref{lumfunc}.

\subsection{Sample completeness}
\label{completeness_lim}
\subsubsection{Detection limit}
\label{detectionlim}

By using the NB1060 image as the detection image (Source Extractor in Dual Image Mode), it is not necessary for an object to be detected in the Y band for it to be included in the sample. In principle therefore, the survey is sensitive to objects with infinitely high equivalent width lines. 

Given that objects are detected solely in the narrow band, the limiting line flux is closely linked to the NB1060 detection limit. The detection limit and detection efficiency are explored by introducing synthetic populations of objects into the images using the IRAF package MKOBJECT. These are then recovered using the detection techniques described in Section \ref{cataloguing}. The detection completeness is the percentage of input objects that are successfully extracted from the images.

As expected for high redshift populations, the real [OII] emitters are not well resolved in the HAWK-I images. The average morphology of the [OII] emitters (measured from the HAWK-I stacked image) is approximated by an exponential disk with scale length r$_s$=2.5$^{\prime \prime}$. Due to the high redshift of the objects and the low resolution, we find the average profile to be representative of the [OII] population as a whole. For this reason, the simulations were limited to a single input profile. 

\begin{figure}
\begin{center}
\epsfig{file=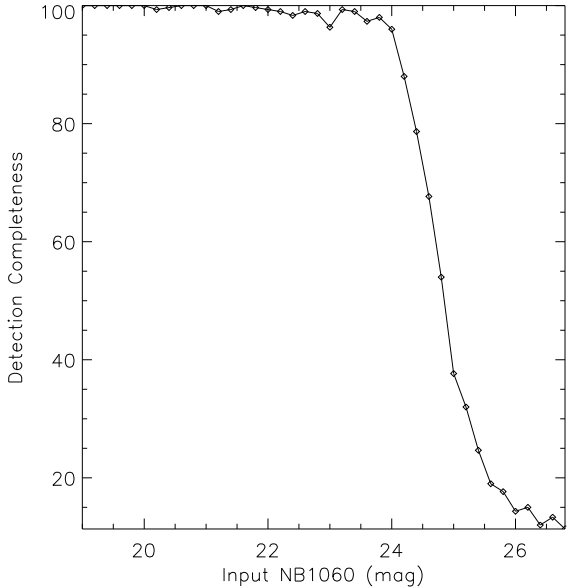,width=1.0\linewidth,clip=}
\end{center}
\caption{Detection completeness as a function of magnitude for synthetic galaxies modelled with the average [OII] emitter profile, an exponential disk with r$_s$=2.5$^{\prime \prime}$.}
\label{detection}
\end{figure}

Batches of 100 objects were introduced into the NB1060 image with continuum magnitudes 18-27 in steps of 0.2 mag. Source Extractor was run on the images using the detection parameters given in Table \ref{separams} and the detection rate was measured. The results of the simulations can be seen in Figure \ref{detection}, which shows how the detection efficiency of the [OII] emitter profile varies with input narrow band magnitude. The Figure indicates that for the \emph{[OII] emitter} profile, the survey is 90\% complete to magnitude 24.2 (cf. 90\% completeness to magnitude 24.4 for point sources (Section \ref{cataloguing})).

\subsubsection{Line flux completion}
\label{lfluxcompleteness}

Whilst sensitive to high equivalent widths, the equivalent width threshold applied in the selection procedure in Section \ref{elgselection} results in objects with EW$_{obs} < 50$\AA  $\;$(equivalent to EW$_{rest}=17.5$\AA) being omitted from the survey.

\begin{figure}
\begin{center}
\epsfig{file=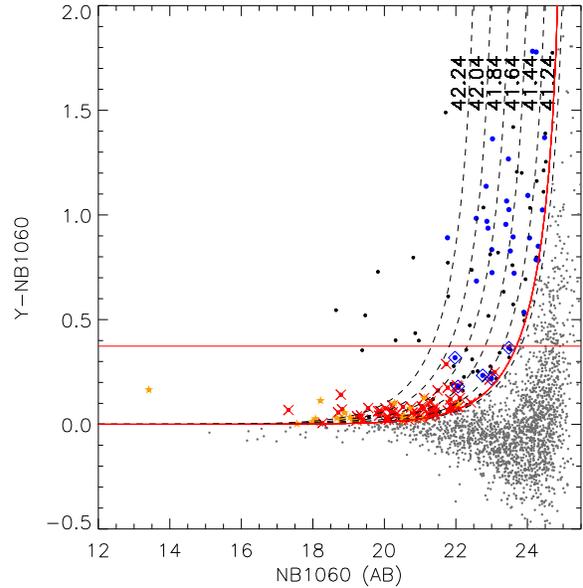,width=1.0\linewidth,clip=}
\end{center}
\caption{Same as Figure \ref{selection}, overlaid with lines of constant line luminosity, labelled logarithmically. The luminosities chosen represent the edges of the bins used to create the fixed-volume luminosity function in Table \ref{lumfunc}. A bin is complete if all the objects between the bin edges are included in the sample. Candidate [OII] emitters are highlighted in blue, stars are indicated by orange asterisks and objects with redshifts placing them out of range of the filter range are shown in red.}
\label{ew_completeness}
\end{figure}

Figure \ref{ew_completeness} shows the selection diagram. Objects above the $\Sigma=3$ selection line are highlighted in bold. ELGs that are potentially missed by the EW$_{obs} > 50$\AA $\;$ threshold lie between the curved and solid red lines. Of the objects falling in this category, we find most of them are either stars (highlighted in orange asterisks) or spectroscopically confirmed redshift interlopers (red crosses). Such a high proportion of confirmed redshift interlopers indicates that the equivalent width threshold chosen in this study is largely robust.

Of the remaining objects, only five have MUSIC photometric redshifts that could lead to them being classified as [OII] emitters, these are highlighted with blue diamonds. We assess the significance of this group of omitted objects by over-plotting lines of constant [OII] line luminosity, representing the edges of the bins of the fixed-volume luminosity function in Table \ref{lumfunc}. For a bin to be complete, all objects between the lines of constant line luminosity corresponding to the edges of the bin must be selected. For example, for the bin centred on $\log(L)=41.74$ erg s$^{-1}$ to be complete, all objects between the lines $\log(L_{[OII]})=41.64$ erg s$^{-1}$ and $\log(L_{[OII]})=41.84$ erg s$^{-1}$ must be selected.

It can be seen that three of the five potential [OII] candidates below the equivalent width threshold have colours placing them in the two lowest luminosity bins in the fixed-volume luminosity function. These bins both lie fainter than the 90\% detection limit of $m_{NB1060}=24.2$ defined in Section \ref{detectionlim}. The two lowest luminosity bins in the study are therefore incomplete due to a combination of detection and selection incompleteness. 

The small number of possible [OII] emitters falling in brighter bins leads us to conclude that we do not lose a significant fraction of emitters using this EW criterion. 

Furthermore, completeness simulations (using the [OII] profile described in Section \ref{detectionlim}) show a recovery rate greater than 90\% for [OII] emitters with Y band continuum magnitudes fainter than magnitude 23 and lines brighter than $\log(L_{line}$)=41.74. This recovery rate rapidly drops off for fainter lines. These results, together with those presented in Figure \ref{ew_completeness}, lead us to conclude that an appropriate completeness limit for the study is the minimum luminosity, $\log(L_{min})=41.74$ erg s$^{-1}$.

\subsection{Schechter fit}
\label{sfit}
The fixed-volume luminosity function is fit with a Schechter function \citep{schechter76} of the form

\begin{equation}
\Phi (L) dL=\phi^* \left (\frac{L}{L^*} \right )^\alpha exp \left (-\frac{L}{L^*} \right ) d \left (\frac{L}{L^*} \right ).
\end{equation}

To compute an appropriate Schechter function for the bright end of the luminosity function, a range of faint end slopes are assumed. 

For comparability, we vary $\alpha$ over the same range as that assumed by Zhu et al. (2009) in their [OII] survey at $0.75<z<1.45$: $\alpha=-1.3\pm$0.2. For comparison, Takahashi et al. (2007) measured the faint end slope of the [OII] luminosity function at \emph{z}=1.2 in two fields finding $\alpha=-1.41_{-0.15}^{+0.16}$ and $\alpha=-1.38_{-0.37}^{+0.40}$.

In Section \ref{lumfunc_evo}, we look at how the luminosity function evolves from \emph{z}=1.2-1.85. We emphasise that the range of $\alpha$ assumed here has a minimal effect as this analysis is restricted to the bright portion of the luminosity function that is well fit, regardless of the assumed faint end slope. 

We fit the [OII] luminosity function at $L  > L_{min}$ with a Schechter function, assuming $\alpha=-1.3\pm 0.2$, using the maximum likelihood parametric fit method (Sandage, Tammann \& Yahil 1979\nocite{sandage79}). The resulting best fit parameters are $\log(L^*)=42.00\pm0.06\;$erg s$^{-1}$ and $\log(\phi^*)=-2.21\pm0.09\;$Mpc$^{-3}$. We note that increasing $L_{min}$ produces fits within these errors but lowering $L_{min}$ quickly departs from them, indicating that $\log(L_{min})=41.74$ erg s$^{-1}$ is an appropriate completeness limit for the survey.

\subsection{Filter correction}
\label{filter_corr}
The fitted fixed-volume luminosity function is scaled to take into account the effects of filter shape. There are two considerations:

{\bf(i)} Flux loss due to filter transmission: ELGs with [OII] lines that fall in the filter wings have lower observed line fluxes due to the poor transmission. When computing the luminosity function, this results in a proportion of line objects systematically moving into fainter bins.

{\bf (ii)} Variation of volume with line strength: Whilst a relatively faint emission line in the filter wings may fall below the detection threshold, a bright line will still be detectable. Brighter lines are therefore detectable over a wider filter width and correspondingly over a wider redshift range and a larger volume. 

Simulations were run to quantify these effects. We considered a volume large enough to fully encompass the filter volume, including the filter wings. The volume was populated with objects with density and flux distributions according to a trial input luminosity function. We assumed the objects were homogeneously distributed with respect to redshift. The filter profile was then used to recover the observed simulated object luminosities. Different values of input $\phi^*$ and L$^*$ were iterated through until the output of the simulation matched the observed [OII] luminosity distribution.

The simulations indicate that the intrinsic observed luminosity function, assuming $\alpha=-1.3\pm0.2$ is best fit with Schechter parameters of $\log(L^*)=42.05\pm0.06\;$erg s$^{-1}$ and $\log(\phi^*)=-2.23\pm0.09\;$Mpc$^{-3}$.

\begin{table}
\centering
\begin{minipage}{\textwidth}
\caption{The Fixed-Volume Luminosity Function}
\begin{tabular}{c c c c c }
\hline
                & \multicolumn{2}{|c|}{Observed Luminosity Function}   \\  
log L([OII])    & log$\phi$                   & Galaxy                  \\
(ergs s$^{-1}$)  & (log L$^{-1}$ Mpc$^{-3}$)    & Counts                 \\
\hline
41.34 & -2.22  & 5  \\
41.54 & -2.22  & 5  \\
41.74 & -2.14  & 6  \\
41.94 & -2.14  & 6  \\
42.14 & -2.44  & 3  \\
42.34 & -2.92  & 1  \\
\hline
\label{lumfunc}
\end{tabular}
\end{minipage}
\end{table}

Figure \ref{lumfunction} shows the observed [OII] luminosity function
found in this study alongside the observed luminosity functions of
equivalent [OII] studies at \emph{z}=1.2. Black triangles show our
binned luminosity function, calculated assuming a constant filter
volume. The luminosity above which the survey is estimated to be
complete ($L_{lim}$) is indicated by the vertical dotted line and the
two bins that are significantly incomplete are highlighted as lower
limits. The best fitting Schechter function to the complete portion of
the luminosity function is indicated with the black dashed
line. Finally, the solid black line is the filter-corrected luminosity
function.  Over-plotted are comparable results from other [OII]
surveys at \emph{z}=1.2. In red is the Schechter fit of Ly et
al. (2007), green points show the binned luminosity functions of
Takahashi et al. (2007) in the COSMOS field (diamonds) and SDF
(asterisks) and the binned luminosity function of Zhu et al. (2009) is
shown in blue. Note that no correction for dust obscuration has been
made to any of the luminosity functions presented in the figure. An
appropriate obscuration correction for our sample is calculated in
Section \ref{dustcorr}.

\section{Evolution of the [OII] Luminosity Function.}
\label{lumfunc_evo}

\begin{figure}
\begin{center}
\epsfig{file=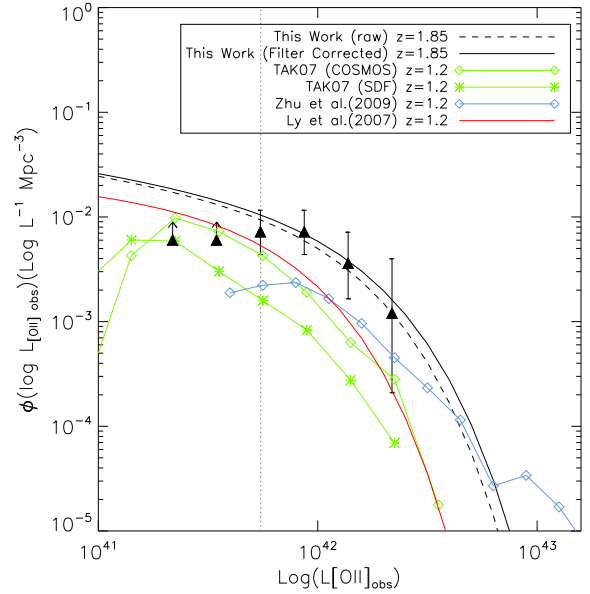,width=1.0\linewidth,clip=}
\end{center}
\caption{Comparison of our observed luminosity function at
  \emph{z}=1.85 with other [OII] luminosity functions from the
  literature at \emph{z}=1.2. Black triangles show the binned
  fixed-volume luminosity function from this work, with the best
  fitting Schechter function in the black dashed line. The final
  observed fit, corrected for filter shape is shown in the solid black
  line. The HAWK-I limiting luminosity is indicated by the vertical
  dotted line. The observed luminosity functions reported by Takahashi
  et al. (2007) are shown in green, that of Zhu et al.(2009) in blue
  and the Schechter fit of Ly et al. (2007) is shown in red. Note that
  none of the luminosity functions shown here have been corrected for
  obscuration.}
\label{lumfunction}
\end{figure}

Figure \ref{lumfunction} suggests the observed [OII] luminosity
function evolves between redshift 1.2 and 1.85. To quantify this
evolution, we compute the number density of [OII] emitters at a range
of redshifts. We concentrate on the integrated number density of
objects in the luminosity range in which our luminosity function is
robust: $\log(L_{[OII]}) > 41.74 $ erg s$^{-1}$.

\begin{figure}
\begin{center}
\epsfig{file=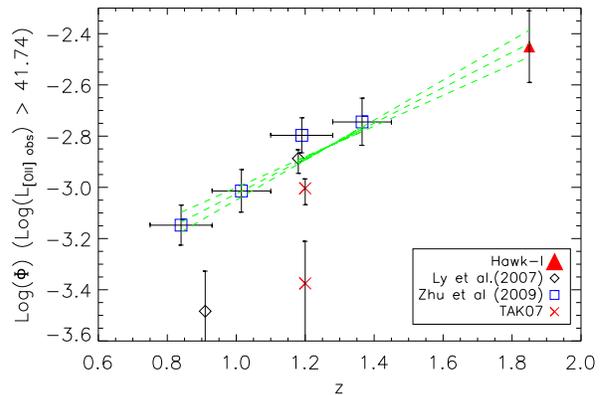,width=1.0\linewidth,clip=}
\end{center}
\caption{Evolution of the observed [OII] luminosity function:
  evolution of the total number of bright objects per Mpc$^{3}$ in the
  luminosity range $\log(L_{[OII] obs}) > 41.74$. The dashed lines
  indicate the best and $\pm \sigma$ fits given by the equation
  $\log(\Phi[\log(L_{[OII] obs})>41.74,\emph{z}])=mz+c$.}
\label{ndevolution}
\end{figure}

We find the number density of objects in the luminosity range {\bf
  $\log(L_{[OII]}) > 41.74$} erg s$^{-1}$ at \emph{z}=1.85 in this
survey is $\log(\rho_{\log(L_{[OII]})>
  41.74})=-2.45\pm0.14\;$Mpc$^{-3}$.  Figure \ref{ndevolution} shows
this result alongside the equivalent number densities of Zhu et
al. (2009) (\emph{z}=0.84, 1.00, 1.19 and 1.35) in blue squares, Ly et
al. (2007) (\emph{z}=0.91 and 1.18) in diamonds and Takahashi et
al. (2007) (\emph{z}=1.2) in red crosses. Our measurement is indicated
by the red triangle.

We fit the points (all but the two outliers) with a line,
$\log(\Phi[\log(L_{[OII]}) > 41.74,\emph{z}])=mz+c$, finding $m=0.69\pm
0.089$ and $c=-3.72\pm 0.11$, this is over-plotted in green on
the Figure, along with the $\pm \sigma$ fits.  We find that the space
density ($\rho$) of bright objects at \emph{z}=1.85 in the luminosity
range $\log(L_{[OII]}) > 41.74$ erg s$^{-1}$ is a factor 2 greater than
the observed space density of [OII] emitters reported at
\emph{z}$\sim$1.4; a comparable increase to that reported between
\emph{z}=1.0 and \emph{z}=1.4.

\section{The Star Formation Rate at \emph{z}=1.85}

Two assumptions have to be made to convert the [OII] luminosity
function found in this work into a star formation rate density: (i)
The [OII] luminosities have to be corrected for internal extinction
and (ii) a calibration has to be assumed between [OII] luminosity and
SFR.

In Section \ref{commonobscorr} we explore the use of a ``common''
obscuration correction to make the results comparable to the
compilation of SFR measurements made by Hopkins (2004). In Section
\ref{dustcorr} we extend the analysis and independently estimate the
extinction correction using 24$\mu$m flux and discuss the appropriate
[OII]/H$\alpha$ ratio.

\subsection{Common obscuration correction}
\label{commonobscorr}

\cite{hop04} provides a compilation of SFR measurements made using a
range of SFR indicators, corrected according to a common framework.
In order to compare the measurement made here to the Hopkins
compilation, we convert our luminosity density to SFR according to
this common framework.

The first step towards computing the SFR is to calculate the total
[OII] luminosity density, $\mathcal{L}[OII]$, at \emph{z}=1.85 by
integrating the Schechter function:

\begin{equation}
\mathcal{L}[OII]= \int^\infty_0 \Phi (L) \ L \ dL=\phi^*L^*\;\Gamma (\alpha+2)
\end{equation}
where $\Gamma$ represents the gamma function.

Hopkins converts the [OII] luminosity density to H$\alpha$ assuming
$\mathcal{L}_{[OII]}/ \mathcal{L}_{H\alpha, obs}$=0.45. The inferred
H$\alpha$ luminosity is then corrected for internal extinction,
assuming A$_{H\alpha}$=1.0 mag. Note that A$_{H\alpha}$=1.0 mag
corresponds to A$_{[OII]}$=1.86 using the O'Donnell (1994) galactic
obscuration curve (R$_v$=3.1). Finally $\mathcal{L}_{H\alpha, corr}$
is converted in to a SFRD using the calibration of Kennicutt (1998):

\begin{equation}
\dot{\rho}_*=7.9\times10^{-42} \rm \mathcal{L}_{H\alpha,corr} \;\; M_\odot yr^{-1}.
\label{hasfrcalib}
\end{equation}

The [OII] Luminosity function derived in this work, corrected for
filter effects (Schechter Parameters: $\alpha=-1.3\pm$0.2,
$\log(L^*)=42.05\pm0.06\;$erg s$^{-1}$ and
$\log(\phi^*)=-2.23\pm0.09\;$Mpc$^{-3}$), implies an [OII] luminosity
density of $\log(\mathcal{L}[OII])=39.93\pm0.08$ erg
s$^{-1}$. Accounting for A$_{H\alpha}$=1mag, this implies a SFRD of
$\dot{\rho}_*=0.38\pm0.06$ M$_\odot$ yr$^{-1} $Mpc$^{-3}$.

\begin{figure}
\begin{center}
\epsfig{file=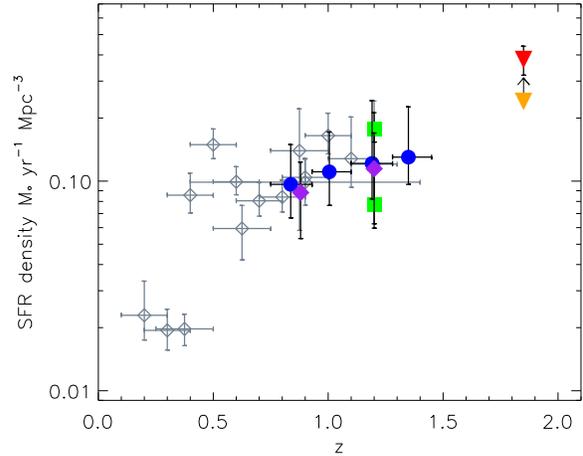,width=\linewidth,clip=}
\end{center}
\caption{A compilation of SFR measurements using the [OII] line,
  assuming a constant obscuration correction, A$_{H\alpha}$=1
  mag. [OII] measurements from the compilation of Hopkins (2004) are
  shown in grey, overlaid with the measurements of Hippelein et
  al. (2003) (purple), Takahashi et al. (2007) (green) and Zhu et
  al. (2009) (blue). The red triangle shows the result from this
    work corrected using the same ``common'' correction as applied to
    the lower redshift points (A$_{H\alpha}$=1 mag). The orange
    triangle shows the same result, this time corrected using
    A$_{[OII]}=0.98$ mag (A$_{H\alpha}=0.52$), the mean obscuration
    of the z=1.85 sample inferred from 24$\mu$m fluxes (see Section
    \ref{dustcorr}).}
\label{oii}
\end{figure}

This result is plotted in red in Figure \ref{oii}, alongside other
SFRD measurements derived from [OII] and corrected using the ``common'' scheme at lower redshift. 
 
Whilst the application of a common obscuration correction is helpful
for comparison with published literature (see Hopkins, 2004), it is an
oversimplification. A more rigorous analysis can be carried out by
measuring the obscuration directly. This is calculated and discussed
in detail in Section \ref{dustcorr}. The final SFRD calculated in
Section \ref{dustcorr} is shown for completeness in orange in Figure
\ref{oii}.

\subsection{Independent measure of A$_{[OII]}$ and [OII]/H$\alpha$}
\label{dustcorr}

\begin{figure*}
\begin{center}
\epsfig{file=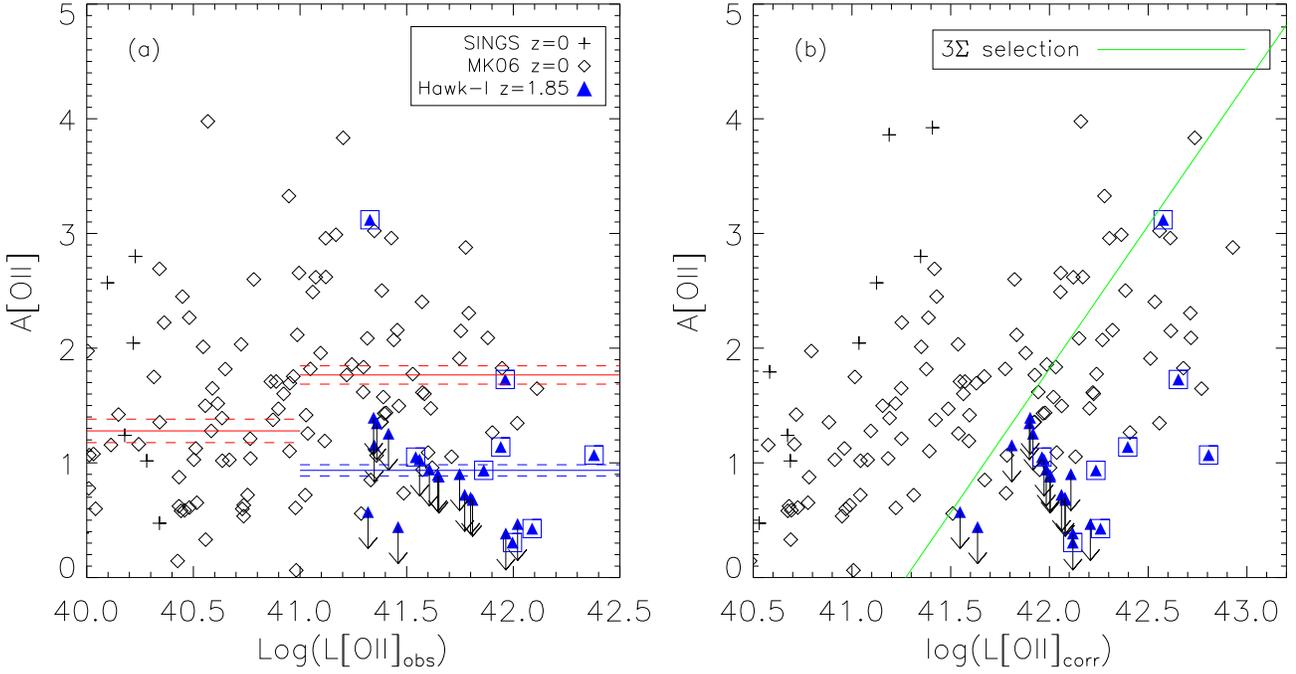,width=1.0\linewidth,clip=}
\end{center}
\caption{(a) [OII] extinction with respect to {\it observed} [OII]
  luminosity for galaxy samples at \emph{z}=0 (MK06 and SINGS samples
  shown in diamonds and crosses respectively) and \emph{z}=1.85
  (HAWK-I shown with blue triangles). 16 of the 24 A$_{[OII]}$ at
  \emph{z}=1.85 were calculated from upper limit measurements of the
  24$\mu$m flux (indicated by the arrows). The eight measurements made
  from positive detections at 24$\mu$m are highlighted with squares
  for clarity. Red and Blue solid lines indicate the mean A$_{[OII]}$
  values of the combined MK06 and SINGS sample and the Hawk-I sample
  respectively. Dotted lines indicate the error on the mean in each
  case. (b) [OII] extinction with respect to {\it obscuration corrected} [OII]
  luminosity, where a correction has been made according to the
  measured obscuration values. The green line shows the maximum
  obscuration with respect to intrinsic luminosity that could be
  selected in our sample.}
\label{loii_aoii}
\end{figure*}

To independently estimate A$_{[OII]}$ for our sample, we look to the
far infrared.  Kennicutt et al. (2009, hereafter K09) developed a set
of empirical calibrations converting [OII] line luminosities into SFRs
using 8, 24$\mu$m or total infra-red (TIR) fluxes as a tracer for the
dust emission. They consider two samples of \emph{z}=0 galaxies; the
SINGS sample of 75 local galaxies with distances less than 30 Mpc,
presented in Kennicutt et al. (2003)\nocite{ken03} and 417 galaxies
from the survey of integrated spectrophotometry described in Moustakas
\& Kennicutt (2006, hereafter MK06)\nocite{mk06}. The samples were
selected to be representative of the wide range of morphologies,
luminosities and dust opacities seen in present day galaxies. The
combined sample includes objects ranging from dwarf irregulars to
giant spirals and IR-luminous galaxies. Full details can be found in
the respective survey papers. For these local galaxy samples, they
find the corrected [OII] luminosity density is given by 
\begin{equation}
L[OII]_{corr}=L[OII]_{obs}+ 0.016L(8 \mu m)
\end{equation}
implying

\begin{equation}
A_{[OII]}=2.5\log \left[1+\frac{0.016L(8 \mu m)_{rest}}{L[OII]_{obs}}\right ].
\end{equation}

For galaxies at the redshift of this survey, $8 \mu$m flux is
redshifted to $\sim 24\mu$m.  Using the 24$\mu$m fluxes from the MUSIC
catalogue, we calculate values of A$_{[OII]}$ for our sample. MUSIC
contains measurements for 24 of the 26 [OII] emitters in the survey,
16 of which are upper limits.

To compare our results with galaxies at \emph{z}=0, we derive the
  [OII] obscuration values for the K09 and MK06 samples using the
  stellar-absorption-corrected H$\alpha$/H$\beta$ ratios quoted in K09
  and MK06. The ratios are converted to [OII] obscurations using the
  same assumptions as detailed in K09; namely we assume an intrinsic
  H$\alpha$/H$\beta$ ratio for Case B recombination of {\it
    I}(H$\alpha$)/{\it I}(H$\beta$)=2.86 (electron temperature T$_e$ =
  10,000 K and density N$_e = 100$ cm$^{-3}$). The observed reddenings
  are then converted to [OII] attenuation values (via H$\alpha$) using
  the O'Donnell (1994)\nocite{odonnell94} extinction law, assuming
  R$_V$=3.1. 

Figure \ref{loii_aoii}(a) shows the distribution of A$_{[OII]}$ with
respect to the uncorrected [OII] luminosity. The \emph{z}=0 SINGS and
MK06 samples are shown in crosses and diamonds respectively. The
\emph{z}=1.85 results from this work are over-plotted in blue
triangles. For the \emph{z}=1.85 objects in our sample, values of
A$_{[OII]}$ derived from upper limits are indicated with arrows and
the eight values of A$_{[OII]}$ derived from positive detections at
24$\mu m$ are highlighted with squares for clarity.

Although the $\emph{z}=0$ measurements of A$_{[OII]}$ have a large
dispersion, we find no evidence for a systematic variation of
A$_{[OII]}$ with observed [OII] luminosity. The mean of
  measurements with $\log(L_{[OII]obs}) > 40.0$ is
  A$_{[OII]}=1.59\pm0.07$ mag with a dispersion of 0.8
  mag. Bifurcating the sample into $40.0 < \log(L_{[OII]obs}) < 41.0$
  and $41.0 < \log(L_{[OII]obs}) < 42.0$ yields mean values of
  A$_{[OII]}=1.42\pm0.1$ and A$_{[OII]}=1.80\pm0.10$ (dispersions of
  0.8 and 0.7 mag) for the fainter and brighter samples respectively,
consistent with no luminosity dependence. The mean values of
each sample are indicated by the red solid lines in Figure
\ref{loii_aoii} and the red dotted lines indicate the error on the
mean in each case.

Our galaxy sample at \emph{z}=1.85 however has systematically lower
levels of A$_{[OII]}$ than the local universe galaxies, with a mean
A$_{[OII]}=0.98\pm0.11$ mag and dispersion of 0.6 mag. This
corresponds to A$_{H\alpha}=0.52$ (O'Donnell (1994) extinction curve
assuming R$_V=3.1$).

Selection effects may bias this result. For any given intrinsic
luminosity, the survey is biased towards selecting objects with low
obscuration: objects with high obscuration values may have (observed)
[OII] fluxes that fall below the selection threshold of the survey. To
investigate the effect this would have on the measured mean
A$_{[OII]}$ value, we re-plotted Figure \ref{loii_aoii}(a), correcting
the observed luminosities for extinction using the measured
obscuration corrections, this is shown in panel (b) of Figure
\ref{loii_aoii}.  The green line shows the maximum detectable value of
A[OII] as a function of corrected-luminosity. The mean obscuration of
the SINGS and MK06 galaxies with observed [OII] fluxes brighter than
the Hawk-I selection threshold is A[OII]$=1.68$, whereas the mean of
all objects with $\log L[OII]_{corr}>41.5=1.90$. Therefore, applying
the Hawk-I selection bias to the \emph{z}=0 sample would result in a
$\sim 12\%$ drop in the mean A[OII] value. It is possible that this
selection limit imposes more than the locally inferred 12\% bias on
the \emph{z}=1.85 sample if a greater proportion of star formation is
dust enshrouded at high redshift. This cannot be ruled out on the
basis of the present study, although it is worth noting that we found
no evidence for a population of bright, dust enshrouded [OII] emitters
(Section \ref{lfluxcompleteness}). Adopting the mean value of
A$_{[OII]}$ at \emph{z}=1.85 (A$_{[OII]}=0.98\pm0.11$,
A$_{H\alpha}=0.52$) yields a SFRD of $0.24\pm0.06 M_\odot$ yr$^{-1}$
Mpc$^{-3}$. This is plotted in orange in Figure \ref{oii}. We have
indicated this result as a lower limit to reflect the fact that a
proportion of dusty [OII] emitters may be omitted from the sample due
to the selection criteria.

In the above analysis we have assumed the widely used [OII]/H$\alpha$
ratio of 0.45 (Kennicutt 1992, 1998). \nocite{ken92} However, local
surveys have suggested that the [OII]/H$\alpha$ ratio is
luminosity-dependent (Jansen et al. 2001)\nocite{jansen01} as well
being dependent on metallicity and obscuration (Kewley et
al. 2004). Hopkins et al. (2003) \nocite{hopkins2003} noted higher
[OII]/H$\alpha$ ratios in higher equivalent width systems. For a
complete sample of 752 SDSS galaxies, they found [OII]/H$\alpha=0.23$
but that this rose to 0.46 if an EW limit of EW(H$\alpha) > 70\; \AA$
was imposed.

Applying the Jansen et al. (2001) empirical relation between the the
rest-frame absolute B band magnitudes and the OII/H$\alpha$ to our galaxy
sample, we find a mean absolute B band magnitude of B$_{abs}$=-20.1
mag ($J=24.5$ mag), corresponding to [OII]/H$\alpha = 0.48$ with an rms
dispersion of 0.1. This is very close to the value (0.45) that we have
assumed, and would result in a SFRD of $0.23\pm0.06 M_\odot$ yr$^{-1}$
Mpc$^{-3}$.

The discussion above highlights that the ``common'' obscuration
correction is an oversimplification of the problem. Upon measuring the
obscuration of the objects we find a mean obscuration of
A$_{[OII]}=0.98$ mag (A$_{H\alpha}=0.52$), rather than A$_{H\alpha}=1$
as assumed in the common framework. However, given that we also find a
proportion of dusty emitters are omitted from the sample, the SFRD
measurement based on the measured obscuration correction is likely to
be a lower limit. It is reasonable to expect that the reality may be
somewhere between the two measurements. Deeper data would enable this
to be explored in more detail.

\section{Summary and Conclusions}

This study has used Science Verification Data from the ESO instrument
HAWK-I to perform a high redshift survey for [OII] emitters in the
GOODS field. The [OII]$\lambda$3727 doublet is of particular interest
to SFRD surveys in that it is visible to \emph{z}$\sim 5$ in the
near-infrared, compared to the \emph{z}$\sim 2.5$ limit of H$\alpha$
surveys. Recent advances in the calibration of the [OII]$\lambda 3727$
doublet (see eg. Kennicutt et al. 2009)\nocite{ken09} have made it
possible to use [OII] with greatly improved precision, facilitating
homogeneous measurements of the SFRD of the universe to be measured
from \emph{z}=0 to \emph{z}=5 for the first time.  At \emph{z}=1.85,
this is the highest redshift [OII] survey to date. We have identified
26 [OII] emitters in a volume of 4138 Mpc$^3$, to a 5$\sigma$ flux
limit of 1.5 $\times$ $\rm 10^{-17} erg \ cm^{-2} s^{-1}$.  Our
findings can be summarised as follows:

\begin{itemize}
\item The observed [OII] luminosity function at \emph{z}=1.85 can be
  fit by a Schechter function with $\log(L^*)=42.05\pm0.06\;$erg
  s$^{-1}$ and $\log(\phi^*)=-2.23\pm0.09\;$Mpc$^{-3}$, assuming
  $\alpha=-1.3\pm$0.2; a representative range of high-\emph{z} faint
  end slopes, including the value reported by Takahashi et al. (2007)
  at \emph{z}=1.2. This range was also assumed by Zhu et al. (2009) in
  their survey of \emph{z}=1.2 [OII] emitters.

\item The space density ($\rho$) of bright ($\rm log(L_{[OII] obs}) >$
  41.74) [OII] emitters at \emph{z}=1.85 is $\log(\rho)=-2.45\pm0.14$
  Mpc$^{-3}$: a factor of 2 greater than the observed space density of
  [OII] emitters reported at \emph{z}$\sim$1.4. This is a comparable
  increase to that reported between \emph{z}=1.0 and \emph{z}=1.4 and
  we find that the [OII] number density evolution of objects in the
  range $\rm L_{[OII] obs} > 41.74$ can be fit by the function,
  $\log(\Phi[\rm \log(L_{[OII]}) > 41.74,\emph{z}])=m\emph{z}+c$,
  finding $m=0.69\pm 0.089$ and $c=-3.72\pm 0.11$.

\item We convert the [OII] fluxes into a SFRD using the ``common''
  extinction correction (A$_{H\alpha}$=1.0) and [OII]/H$\alpha$ ratio
  ([OII]/H$\alpha$=0.45) employed by Hopkins (2004), finding a SFRD at
  \emph{z}=1.85 of $\dot{\rho}_*=0.38\pm0.06\;$M$_\odot$
  yr$^{-1}$. When compared to other reported values of the SFRD,
  calculated using the [OII] emission line and the ``common''
  conversion, our work suggests a three fold increase in the SFRD
  between \emph{z}=1.4 and \emph{z}=1.85.

\item We independently estimate A$_{[OII]}$ for each object
  using rest frame 8$\mu$m flux (observed 24$\mu$m) and the empirical
  calibrations of Kennicutt et al. (2009). The results indicate that
  the [OII] emitters we detect contain low levels of dust - the mean
  extinction of the sample being A$_{[OII]}=0.98\pm0.11$ with a
  dispersion of 0.6 (equivalent to A$_{H\alpha}=0.52$). This is
  almost half the value measured at \emph{z}=0.  

\item Possible explanations of the low dust content are: (i)
    Selection Effects: The survey could be missing bright, dusty [OII]
    emitters due to the equivalent width threshold applied to the
    sample. This is unlikely as we find no evidence to indicate the
    survey misses bright objects with low-equivalent widths (Section
    \ref{lfluxcompleteness}).  We do find however that selecting
    objects above a fixed {\it observed} [OII] flux threshold may
    account for some of the difference between our A[OII] measurement
    at \emph{z}=1.85 with respect to \emph{z}=0. This is because
    objects with high levels of obscuration fall below the {\it
      observed} [OII] flux threshold leading to an underestimation of
    the true value. We estimate this bias results in an
    underestimation of the true mean by at least $\sim 12\%$. However,
    without deeper data, it is unclear whether the bias alone could
    account for the full ($\sim50\%$) reduction in dust obscuration
    which is seen between the \emph{z}=1.85 and unbiased \emph{z}=0
    samples. (ii) Cosmic variance: the survey is relatively small
  ($\sim 4000\;$Mpc$^{-3}$) and therefore may not be representative of
  a typical region at \emph{z}=1.85. (iii) Reddening Law: The K09
  empirical calibrations used to derive A$_{[OII]}$ at \emph{z}=1.85
  in this work were measured at \emph{z}=0. Different reddening laws
  may apply at \emph{z}=1.85. (iv) Lower dust content at \emph{z}=1.85:
  the tight dispersion supports the idea that there is genuine,
  measurable difference between the amount of dust in high and low
  redshift [OII] emitters. This would fit in with previous reports of
  low dust content in high-redshift galaxies (see e.g. Bunker et
  al. 2010, Ho et al. 2010 and references therein). \nocite{bunker10}
  \nocite{ho10}

\item Incorporating the high-redshift value of A$_{[OII]}=0.98$ into
  the SFR estimate yields a final SFRD of $0.24\pm0.06 M_\odot$
  yr$^{-1}$ Mpc$^{-3}$. This is a lower limit on the star formation
  density at \emph{z}=1.85, given that a proportion of dusty emitters
  are omitted from the sample due to the bias noted above. This is the
  first result tracing the SFRD to \emph{z}=1.85 using [OII]. It is
  in agreement with the UV SFRD measurements of Reddy et al. (2008)
  who found $\dot{\rho}_* = 0.21$ (with an error $\sim 20\%$) at
  $z\sim 2$.  \nocite{reddy2008}

\end{itemize}

\section{Acknowledgements}
We thank the anonymous referee for a careful reading of this
manuscript and for providing suggestions that have improved it. We
gratefully acknowledge Manda Banerji for producing the galaxy
colour-evolution tracks plotted in Figure \ref{tracks} and also Paul
Hewett for providing stellar evolution tracks used to calibrate the
data.

\bibliography{Bayliss_goods_oii_paper}{}
\bibliographystyle{mn2e}

\end{document}